\pdfoutput=1
\documentclass{JINST}

\title{Development of a modular and scalable data acquisition system for calorimeters at a linear collider}

\author{M.J.~Goodrick$^a$, L.B.A.~Hommels$^a$, R. Shaw$^a$, D.R.~Ward$^a$,
D.S.~Bailey$^b$\thanks{Now at the The Mathworks (UK) Ltd.}~, M. Kelly$^b$\thanks{Now at Curtiss-Wright Controls Embedded Computing, High Wycombe.}~, 
V.~Boisvert$^c$, B. Green$^c$, M.G. Green$^c$, A. Misiejuk$^c$, T. Wu$^c$,
V.~Bartsch$^d$\thanks{Now at the University of Sussex.}, M.~Postranecky$^d$, M.Warren$^d$ and M. Wing$^d$\thanks{Corresponding
author.}\\
\llap{$^a$}University of Cambridge,\\
  Department of Physics, Cavendish Laboratory, J.J. Thomson Avenue, Camridge, CB3 0HE, UK\\
\llap{$^b$}University of Manchester,\\
  The School of Physics and Astronomy, Oxford Road, Manchester, M13 9PL, UK\\
\llap{$^c$}Royal Holloway, University of London,\\
  Department of Physics, Egham, Surrey, TW20 0EX, UK\\
\llap{$^d$}UCL,\\
  Department of Physics and Astronomy, Gower Street, London, WC1E 6BT, UK\\
  E-mail: \email{M.Wing@ucl.ac.uk}}

\abstract{
A data acquisition (DAQ) system has been developed which will read out and control calorimeters 
serving as prototype systems for a future detector at an electron-positron linear collider.  This is a 
modular, flexible and scalable DAQ system  in which the hardware and signals are standards-based, using 
FPGAs and serial links.  The idea of a backplaneless system was also pursued with a commercial 
development board housed in a PC and a chain of concentrator cards between it and the detector forming the 
basis of the system.  As well as describing the concept and performance of the system, its merits and 
disadvantages are discussed.
}


\begin{document}

\section{Introduction and concept}

Calorimeters planned by the CALICE collaboration~\cite{calice2} for a future electron-positron linear collider will 
be highly granular with cell sizes as low as $5 \times 5$\,mm$^2$~\cite{cal}.  This fine granularity is needed in order to 
maximise the performance of using particle flow algorithms~\cite{pfa} to reconstruct the final state and in particular to 
achieve the best possible jet energy resolution.  Such segmentation results in a calorimeter with about 100 million cells 
which need to be read out.  Zero suppression in the front-end electronics will reduce the number of cells to be read out 
per event and hence reduce the data rates and number of components needed for a data acquisition (DAQ) 
system.  Additionally a series of data concentrators at various stages can be applied, leading to a funnel-like DAQ system.  
A motivation of the structure of the DAQ system presented here was to perform R\&D~\cite{DAQ-proposal} and design 
work so that it could in principle be scaled to the needs of a final detector at a linear collider, whilst also being used by the 
technical CALICE prototypes~\cite{calice2} being built as part of the EUDET project.  An important 
requirement of the DAQ system was a triggerless acquisition in which the downstream elements should accept anything 
they receive. 

A strong underpinning thread for the design of the DAQ system was to make use of commercial and standards-based 
components, such as FPGAs and existing serial-link hardware, and to identify any problems with this approach.  
Therefore the system should be easily upgradable, both in terms of acquiring new components at competitive prices, 
and easily scalable.  The concept of moving towards a "backplaneless" readout and using PCs instead of a crate-based 
system is also 
pursued.  These goals had to be balanced with the need to deliver a DAQ system to be used by the CALICE technical 
prototype calorimeters in a future test-beam programme.  The DAQ system must be flexible enough to read out the 
electromagnetic calorimeter (ECAL), the analogue hadronic calorimeter (AHCAL), and the digital hadronic calorimeter 
(DHCAL), which are all under development within the CALICE Collaboration.  Indeed it is sufficiently generic such that 
it could be used by many different sub-detectors at a linear collider or elsewhere.  

The article is organised as follows.  The overall architecture of the DAQ system is discussed in Section~\ref{sec:overall}, 
with a detailed description of each component and its functionality given in Section~\ref{sec:comp}.  In 
Section~\ref{sec:firmware}, an overview of the firmware and links is given.  The system performance in terms of 
throughput and efficiency is presented in Section~\ref{sec:perf}.  In Section~\ref{sec:discuss}, the general philosophy 
of the system is discussed with e.g. commercial off-the-shelf and bespoke equipment contrasted.  Finally, a summary 
is given in Section~\ref{sec:summary}.

\section{Overall architecture of the DAQ system}
\label{sec:overall}

\begin{figure}[htbp]
 \begin{center}
  \includegraphics[width=0.95\textwidth]{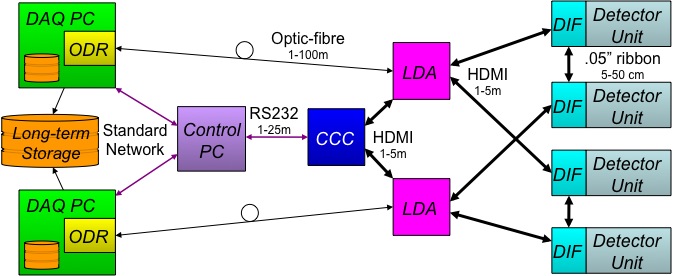}
   \caption{\label{fig:overview} Overview of the DAQ system showing the individual components and links between them.  The 
   components are the detector interface (DIF) card, the link data aggregator (LDA), the clock and control card (CCC) and the 
   off-detector receiver (ODR) housed in a DAQ PC.  For details of their functionality, see text.
   }    
 \end{center}
\end{figure}

A block diagram of the overall structure of the DAQ system is shown in Fig.~\ref{fig:overview}.  This shows the modular 
structure of the system and the different levels of concentration as data passes from the detector unit (in this case a 
calorimeter layer) to mass storage.  Each component is discussed briefly below to give an overall picture of the DAQ 
system, with detailed functionality covered in Sections~\ref{sec:comp} and~\ref{sec:firmware} .  The system is 
bi-directional so although the description below is for the transport of data collected by the detector and sent to mass 
storage, it equally applies to e.g. control data being sent in the reverse direction from a control PC to the detector units.

A detector unit will typically be one layer of a calorimeter module in which data is initially accumulated by ASICs before 
being passed along to the end of the layer.  A layer is a set of ASICs collected together for readout and is given by the full 
length of the calorimeter and a width dependent on the calorimeter type.  At the end of the calorimeter layer, the 
detector interface (DIF) card 
aggregates the data from the ASICs.  
Due to the 
different calorimeter designs involved, the DIFs are mechanically and electrically different for each calorimeter type.  Hence 
there is a different DIF design for each of the ECAL, AHCAL and DHCAL.  As the rest of the system after the DIF is 
calorimeter independent, the data are converted into a common format by unifying certain aspects of the firmware.  In 
this way, the board could be viewed as being partly calorimeter-specific and partly generic.  This also implies that not 
just calorimeters but any sub-detector, within the constraints of data transfer (see Section~\ref{sec:perf}), could be read 
out with the DAQ system discussed here, given the appropriate detector-specific DIF.

A DIF is connected to a link data aggregator (LDA) which is essentially a concentrator card taking 10 DIFs as input and 
passing the data on one link to the next stage of the DAQ.  The link between the DIFs and LDAs is via HDMI cables of around 
1\,m in length as the LDA is envisaged to be housed at the end of the detector unit close to the DIFs.  As HDMI is a commercial 
standard for consumer electronics, high-bandwidth connections are achieved at low cost.  The limitation of the 
LDA connecting to 10 DIFs is due to the space constraints of 10 HDMI connectors on the board and the amount of data the 
FPGA on the LDA can handle.  A future system could try and further optimise the number of links in order to reduce the 
number of components.  

The aggregated data from the LDA is then passed along an optical fibre via gigabit Ethernet to an off-detector receiver 
(ODR), a commercially available development board, housed in a PC.  The ODR again acts as a concentrator card, accepting 
data from four LDAs.  The data is then passed from the PC to a RAID storage array.  The ODR, PC and data storage 
will be in the counting room, with fibres of the order of tens of metres connecting the ODRs to the LDAs.  The long-term 
storage will not be in the DAQ PC and is to be defined.

To ensure the electronics captures data from actual events, all LDAs, and hence all components on the detector, are 
synchronised.  A fast  clock is distributed with synchronisation signals which allow reconstruction of the accelerator 
clock.  As the CCC will be used in different beamlines, it can accept a range of frequencies (50--150\,MHz) 
via LVDS, TTL or NIM inputs or provide a 50\,MHz standalone clock.  
The value of 50 MHz was chosen as it can also be used for the data links and is sufficient for the bandwidth 
required (see below). 
Signals for the start and end of data taking as well as synchronisation and busy signals are required.  
This is achieved with a clock 
and control card (CCC) which fans out the machine clock and fast signals to all LDAs via an HDMI link.  The CCC will 
be situated in a crate (although it has a separate power supply) close to the LDAs in the detector area.  A control PC 
hosts a run control and software to interface to the 
hardware and control the passage of data to and from the detector.  The use of an RS232 link to the CCC allows the 
control PC to be placed outside the detector area.

The principal requirements of the DAQ system developed here are given by the need to read out the three different 
prototype calorimeter modules, the ECAL, AHCAL and DHCAL.  Use of the DAQ system in beam tests is expected 
in an initial phase at the end of 2011 and during 2012.  A calorimeter module has 30--50 layers, depending on the type 
of calorimeter.  The number of channels per layer and hence per calorimeter type differs, however the data is 
aggregated such that the rate from each DIF is similar.  As an example, in the case of the ECAL, the prototype module 
will be 30\,layers deep with a total of about 38\,000 channels split into 64 channels per ASIC for read out.  The practical 
data limit assumes the readout of ASICs on a detector slab are organised in four parallel daisy-chains, each running      
at a 5\,MHz clock, giving a maximum input data rate for a DIF of 20\,Mbits/s.  It should be noted that this rate is only        
sustained as long as there is data available in the ASICs.  Hence, the LDA--DIF serial link has a bandwidth set to 
50\,Mbits/s with 8bit/10bit encoding, giving a theoretical transfer limit of 40\,Mbits/s, above the required value.  During 
beam tests, the ASICs will auto-trigger, accepting and passing on any data to the DAQ system.  During cosmic or 
laboratory tests in which there is an external trigger, the LDA and on-detector components need to be synchronised 
to below 5\,ns, as dictated by the ASIC design~\cite{calice2}.   

\section{Functionality of each component}
\label{sec:comp}

\subsection{The detector interface card}
\label{sec:dif}

\begin{figure}[htbp]
 \begin{center}
  \includegraphics[width=0.75\textwidth]{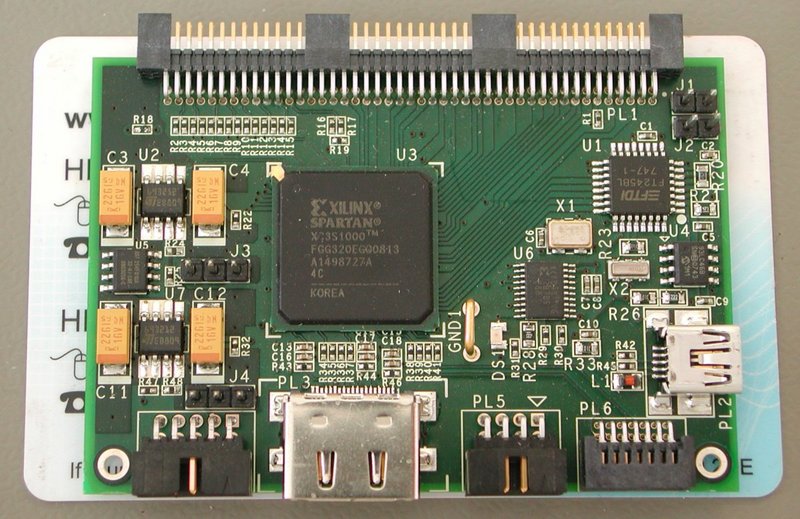}
   \caption{\label{fig:dif} Picture of an ECAL DIF card overlaying a credit card to give an impression of size.  The HDMI 
   link to an LDA is at the bottom immediately below the FPGA.  The connector shown on the bottom of the board at 
   the far right links a DIF to another DIF via a 0.05\,inch ribbon.  The USB connector on the right-hand side of the 
   board is used for standalone and debugging.
   }    
 \end{center}
\end{figure}

Although different in detail, the different calorimeter DIFs are in principle similar with relevant groups designing and 
producing their calorimeter-specific electronics in consultation with the other groups.  An example, the ECAL DIF, is 
shown in Fig.~\ref{fig:dif}; it has dimensions of about $7 \times 5$\,cm$^2$.  The top of the board connects to the 
detector electronics.   The connection with the DAQ is established by a LDA--DIF link, running over AC-coupled 
LVDS differential pairs on a standard HDMI cable and connector shown at the bottom of the picture.  To provide 
extra redundancy in case of a DIF--LDA link failure, a given DIF can be connected to a neighbouring DIF and the 
data transferred via this path until the DIF--LDA link is recovered.

The functionality of the DIF is as follows:

\begin{itemize}

\item to receive (or generate) timing control from the CCC, directly or via an LDA, and distribute to the detector ASICs;

\item to receive, decode and store blocks of event data from the detector ASICs and transfer to an LDA;

\item to receive and decode incoming commands from an LDA (or the USB connection) and assert corresponding 
signals;

\item to manage power dissipation by setting the ASICs to low-power mode between bunch trains as they are inside 
the detector and so receive no dedicated cooling.  The DIFs and all components off the detector take advantage of the 
inter-train period to read out data;

\item to control the DIF--DIF redundancy connection.

\end{itemize}

\subsection{The link data aggregator card}

\begin{figure}[htbp]
 \begin{center}
  \includegraphics[width=1.0\textwidth]{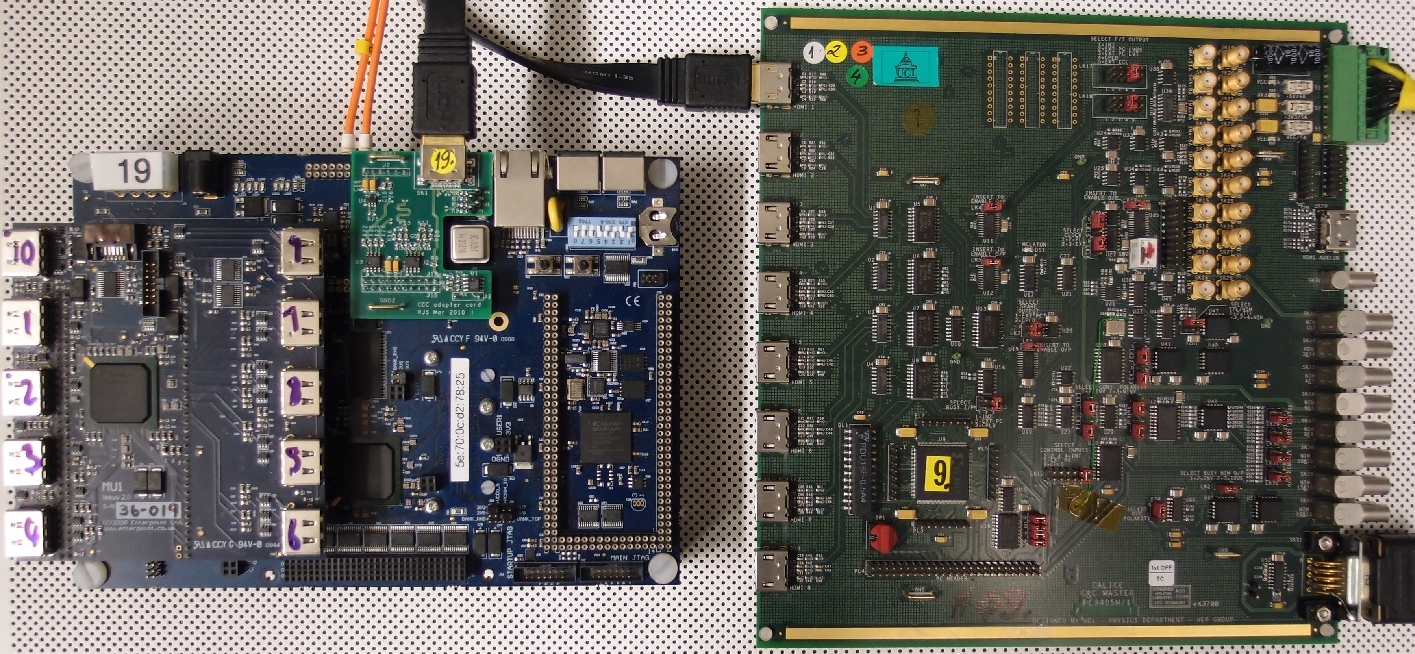}
   \caption{\label{fig:lda-ccc} Picture of the (left) LDA consisting of a Mulldonoch\,2 baseboard, a 10-link HDMI board 
   (from Enterpoint Ltd~\cite{enterpoint}) and an adaptor board with a link, via the HDMI cable, to the (right) clock and 
   control card.  The Ethernet add-on board~\cite{enterpoint} for the LDA is not visible as it is attached to the underside 
   of the baseboard with the optical cable plugged into it.
   }    
 \end{center}
\end{figure}

The LDA mechanically consists of four boards, three of which have been procured from a commercial 
company~\cite{enterpoint} and one designed in-house (the link to the CCC).  The system is shown on the left in 
Fig.~\ref{fig:lda-ccc}.  The 10 HDMI connectors, to link to the DIFs, are contained on a separate board which is 
attached to the main FPGA (Xilinx Spartan-3 XC3S2000) baseboard.  The 1\,Gbit/s Ethernet link to the ODR is 
also on a separate board, attached to the underside of the baseboard (not seen on this photograph).  The link 
(small add-on board) to the CCC is also shown.

The functionality of the LDA is as follows:

\begin{itemize}

\item to receive and multiplex 10:1 data from the DIFs to be subsequently sent to the ODR;

\item to receive and combine the busy signals from 10 DIFs;

\item to fan-out 1:10 fast signals from the CCC to the DIFs.

\end{itemize}

\subsection{The clock and control card}

The requirements on the clock are a low jitter (measured to be about 120\,ps) and a fixed latency between the 
machine clock and the clock in the DIFs.  Hence, the CCC is a custom-made board taking into account the 
needs of the different calorimeters to which it provides signals.  It is shown on the right in Fig.~\ref{fig:lda-ccc} 
and has dimensions $234 \times 220$\,mm$^2$.  The board contains various connectors with 8 HDMI 
connectors the main fan-outs to 8 LDAs.  As a common link specification is used (see Fig.~\ref{fig:firmware}), a 
CCC can connect to a DIF for standalone operation.  If a small simple DAQ system (e.g. via USB) is used, 
the CCC can be used as a common clock source for up to 8 DIFs.

The functionality of the CCC is as follows:

\begin{itemize}

\item fanning out of common machine clock to all detectors;

\item fanning out a standalone clock, if needed, to all detectors;

\item synchronisation of all slow clocks on all DIFs;

\item transfer of all asynchronous triggers as fast as possible;

\item receive busy command from DIFs (via LDAs) to stop the data acquisition.

\end{itemize}

To send the busy on AC-coupled lines, the signal was encoded as a clock.  The trigger is a pulse and could therefore 
be transferred over AC-coupled lines.

\subsection{The off-detector receiver and DAQ PC}
\label{sec:odr}

\begin{figure}[htbp]
 \begin{center}
  \includegraphics[width=0.75\textwidth]{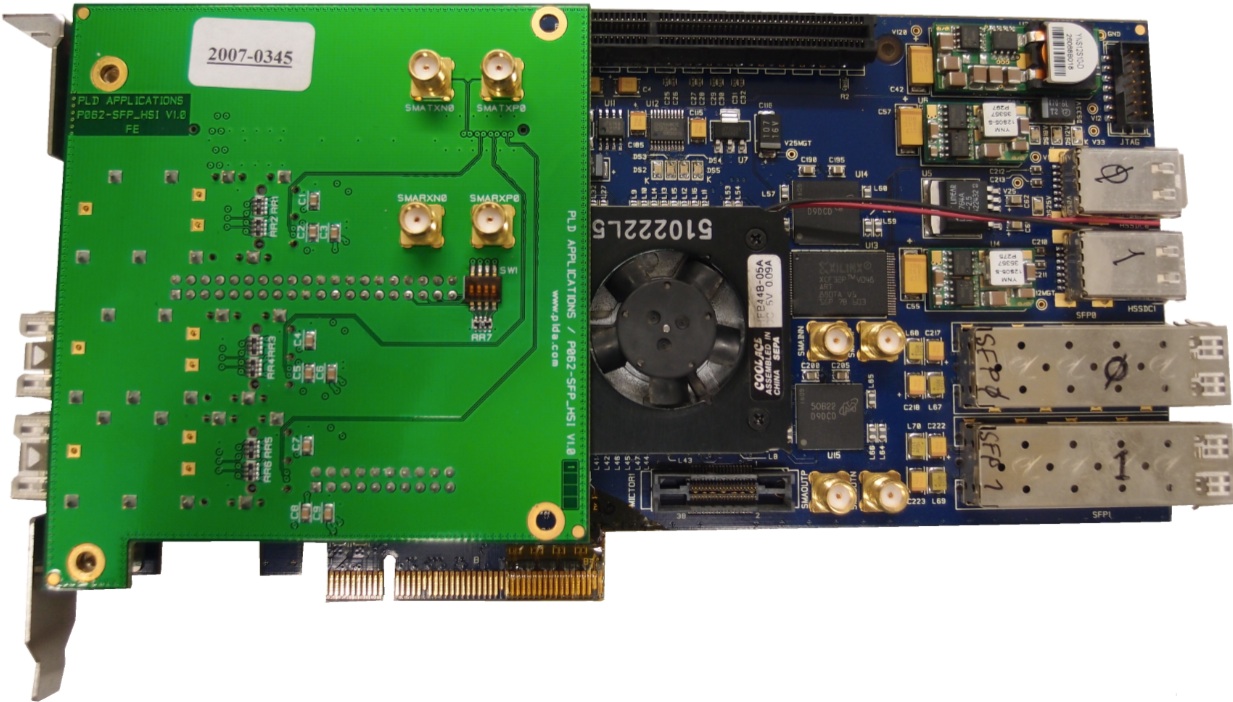}
   \caption{\label{fig:odr} Picture of the ODR (from PLDA~\cite{plda}), showing the main board with its two SFP 
   fibre connectors, bottom right (the two HSSDC2 connectors above them are not used in this DAQ system).  The  
   add-on board provides two additional SFP connectors.
   }    
 \end{center}
\end{figure}

The ODR is a commercially-available~\cite{plda} FPGA development board, shown in Fig.~\ref{fig:odr}.  Its main 
features are a powerful FPGA (Xilinx Virtex-4 FX100), four SFP fibre connections to receive and send data, fast 
write to disc via an eight-lane PCI\,Express bus  and significant memory storage for buffering data.  The powerful 
FPGA which could allow for extra processing and data reduction as well as the large memory and ability to store 
data until it can be read out are the two features which distinguish the ODR from a conventional network card.  
However, in low-bandwidth laboratory tests, a commercial network card can be used as data is transferred using 
raw Ethernet protocol.  The ODR is controlled by Linux drivers and housed in a standard PC.

The functionality of the ODR is as follows:

\begin{itemize}

\item to receive data from four LDAs, potentially buffer and write to disc;

\item send control data to the detectors.

\end{itemize}

The DAQ PC is housed in a case which can be placed in a shelf or rack-mounted.  This contains a CPU with 
four cores, 3\,GBytes of memory and four PCI\,Express slots available for a maximum of four ODRs.  A RAID 
array of four discs configured as a RAID 5 array is attached to the DAQ PC.

The performance of the ODR was determined:  the rates of writing to memory for four PCI Express lanes 
and the rates of writing to the RAID storage array for different methods were measured.  The maximum 
transfer to memory was measured to be about 700\,MBytes/s, below the link rate of 
10\,Gbits/s due to overheads of the protocols used to implement PCI\,Express (such as 8b/10b encoding) 
and additional protocol overheads due to management of the ODR.  The rate of writing to disc 
is significantly lower and depends on the method used.  A write method based on the scatter-gather 
technique\footnote{The scatter-gather technique provides data transfers from one non-contiguous block 
or memory to another by means of a series of smaller contiguous-block transfers.}, rather than writing a single event 
fragment to a single file, yielded the highest 
transfer rates of up to 280\,MBytes/s.  This value represents the practical limit of writing to the RAID array of this 
configuration.

\section{Firmware and links}
\label{sec:firmware}

\begin{figure}[htbp]
 \begin{center}
  \includegraphics[width=1.\textwidth]{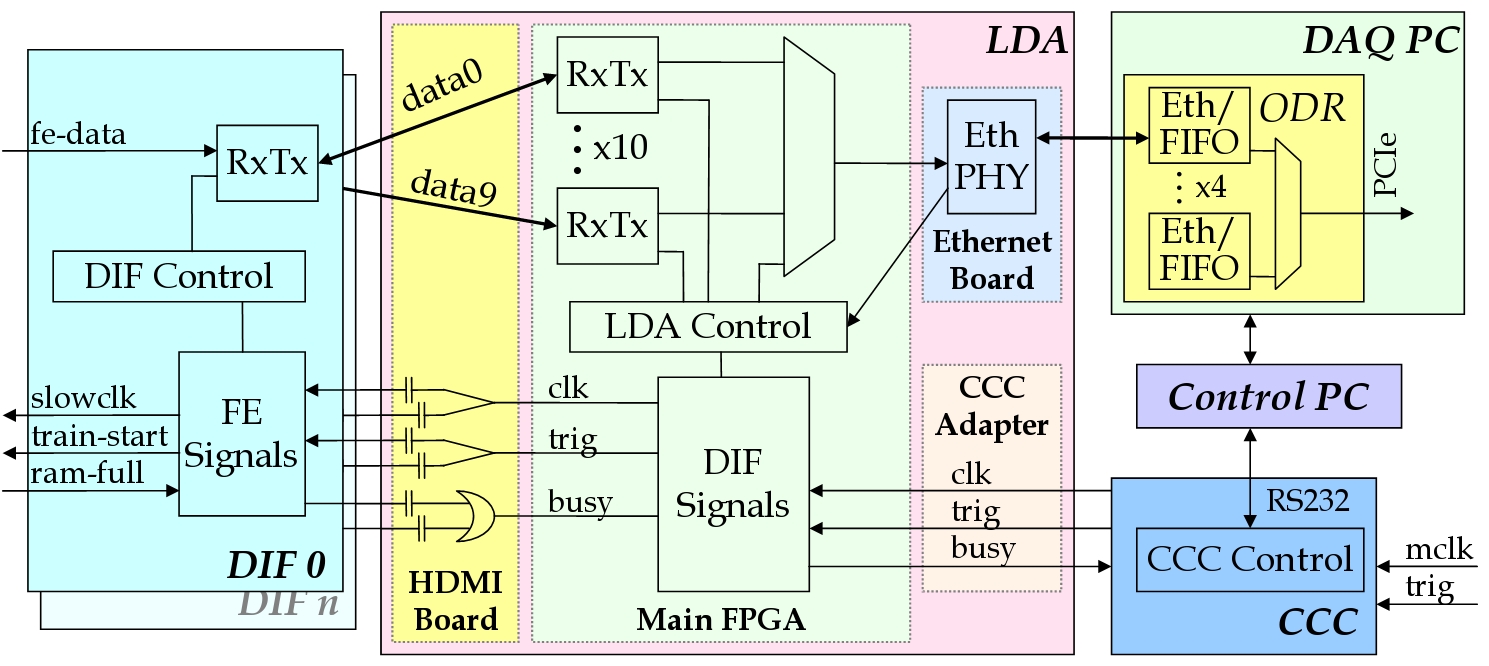}
   \caption{\label{fig:firmware} Block diagram showing an overview of the firmware structure and interconnects.}    
 \end{center}
\end{figure}

Descriptions of the firmware and links are given below and an overview shown in 
Fig.~\ref{fig:firmware}.  The full structure and code is freely available~\cite{daq-fw}.

Packet-based communication is used for the DAQ system.  All control data is sent in packets from the 
DAQ PC, via the ODR, to the LDA which will forward them to the DIF as needed.  Control 
packets can be targeted at a specific LDA or DIF or broadcast.  Communication uses 
Ethernet protocol (with each LDA having a unique MAC address) such that it can also function on 
a generic network.  The CCC is controlled by an RS232 connection from a control PC and 
has various other connectors to receive e.g. a machine clock or external trigger.                                                                    

Data readout is structured hierarchically.  Data flows from the detector ASIC via a DIF, LDA and 
ODR, to the DAQ PC memory and ultimately to hard disc storage.  At each stage more streams 
are aggregated, requiring higher bandwidths. Data are packaged in blocks of up to 1.5\,kBytes to 
facilitate the use of non-Jumbo Ethernet frames.  First-in first-out (FIFO) memories are used to gather 
a full packet before it is forwarded onto the next stage. Additional wrappers and headers are also added to 
improve the traceability of the data.

Signalling between the DIF and LDA makes use of 5 differential LVDS pairs in an HDMI cable.  To 
allow the DIF to be AC-coupled to the LDAs, 8b/10b encoded signalling is used for data.  Also 
transmitted to the DIF is a clock and asynchronous trigger signal.  The DIF sends a busy to the LDA 
using a clock for assert.  The CCC--LDA link has the same HDMI cable and signal specification 
as the LDA--DIF link, except the data lines are unused; this allows the CCC to be plugged directly into 
a DIF.  

\section{System performance}
\label{sec:perf}

\begin{figure}[htbp]
 \begin{center}
  \includegraphics[width=0.85\textwidth]{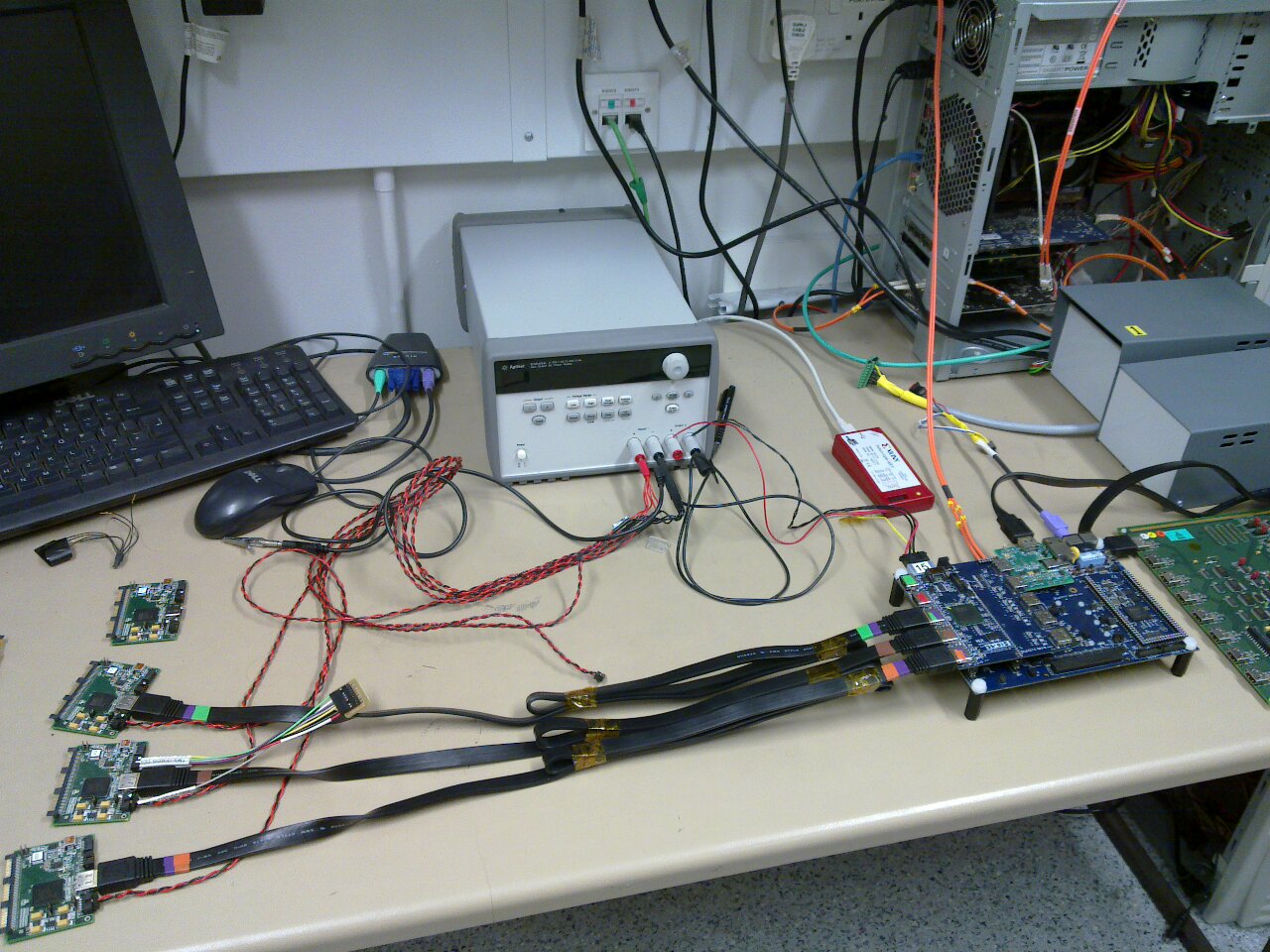}
   \caption{\label{fig:system} Picture of an example test set-up:  three DIFs (bottom left) are connected via 
   HDMI cables to an LDA.  The LDA Ethernet board is connected via the optical fibre to the ODR 
   housed in the DAQ PC (top right).  The CCC (centre far right) provides the clock and trigger via an HDMI 
   cable connected to the adaptor board attached to the LDA.
   }    
 \end{center}
\end{figure}

The system was put together initially for integration of all parts, then for debugging and finally for tests of the 
performance.  All tests described in this section were performed on a laboratory test-bench using simulated 
data; final tests will be performed with future prototype modules.  For visual clarity, a simple example test 
set-up is shown in Fig.~\ref{fig:system}.  In general, a full compliment of 10 DIFs were used and connected 
to an LDA.  Test packets of data were sent via a simple command from a PC via the ODR through the LDA 
and to the DIFs.  The DIFs received this data and sent back encoded information such that the DAQ PC 
determined that a given packet was received by the DIF.  By varying the amount of data sent, the throughput 
and efficiency of the system were measured.

The transfer of signals over AC-coupled LVDS lines presented challenges.  For example a clock is used to 
indicate BUSY and undriven/floating to indicate NOT-BUSY.  This relies on the LVDS receivers defaulting 
to a known output state when their inputs are undriven (i.e. input pins at near equal potential). As the default 
state differs between LDA and DIF receivers, logic levels were inverted where appropriate.                                                                                               

As stated in Section~\ref{sec:overall}, the theoretical limit of the DIF--LDA link is 40\,Mbits/s, twice the practical limit 
of 20\,Mbits/s.  Using 10 DIFs connected to one LDA, a sustained transfer rate of 28\,Mbits/s for each link was 
measured, which although 70\% of the theoretical limit, is comfortably above (40\% higher than) the practical limit 
of 20\,Mbits/s.  The firmware developed was a compromise between simplicity and ease of application and peak 
performance.  Further optimisation to increase the transfer rate was not done as this was deemed surplus to 
requirements.

After the LDA combined data from all 10 DIFs, this gave a data rate on the LDA--ODR link of 280\,Mbits/s, comfortably 
below the maximum link speed of 1\,Gbit/s.  All DIF--LDA links and the LDA--ODR link ran efficiently with no data loss 
at rates up to those given above.

From this it can be seen that the highest data rate expected of 20\,Mbits/s for a DIF--LDA link can be handled by all 
further links and components even if all DIFs were running at this rate.  The subsequent rate for the 
LDA--ODR link would be 200\,Mbits/s which for four LDAs amount to a rate of 800\,Mbits/s received by each ODR.  
Assuming two ODRs in a DAQ PC, this equates to a rate of 200\,MBytes/s, below the measured limit (see 
Section~\ref{sec:odr}) of writing to disc of 280\,MBytes/s.

\section{Discussion}
\label{sec:discuss}

The approach to designing the DAQ system outlined here was to look at new technologies and, to some extent, 
to do R\&D on DAQ systems in general and look at ways in which they could be improved.  This was all 
balanced by the need to provide a DAQ system for use in beam tests by future calorimeter prototypes designed 
for an electron-positron linear collider.  Some of the lessons learnt from new approaches can only be detailed 
after extensive long-term testing such as deployment in a beam test and used for several years.  However, 
even after the development, as is the case here, contrasting certain aspects of the R\&D with more traditional 
methods is possible and given here.

The use of a backplaneless system was motivated by the cheapness of PCs and the use of commercial 
standards in the electronics and networks which were well served by PCs rather than crates.  At the time of 
design of the system, the telecommunications crate architecture (TCA) standard~\cite{tca} was in its infancy and not 
considered.  Desktop PCs can be obtained at low cost, but are less reliable than the more expensive 
crate-based solutions often used for large detectors.  A crate system such as TCA presents a reasonable 
solution with the reliability and ease of access associated with crate systems.  However, given the system 
here is designed using such standards as PCI\,Express and FPGAs, much of it can be ported to a TCA 
system should this route be chosen.

The use of commercial standards is one of the major thrusts of this DAQ system and as said above it should 
ensure portability to other systems using similar standards.  In principle, the system should also be easily 
upgradable if  a company produces e.g.  a new FPGA family or as higher speed Ethernet components become 
available.  The 
advantage of using commercially-available boards is less clear due to boards not always having all the 
required functionality.  It is more important to use individual components which are commercial and 
standards-based rather than the board as a whole.

\section{Summary}
\label{sec:summary}
In summary a data acquisition system has been developed which will be used to read out future calorimeter prototypes 
for an electron-positron linear collider.  The system uses commercial and standards-based components where possible and 
has a modular structure of data concentration at various layers.  Tests have been made which show that the performance 
of the system is well above that required for the calorimeters.  Given the generic nature of the system, it could also be used 
to read out other detectors at other experiments with data rates that are commensurate with those measured here.

\acknowledgments
We acknowledge the rest of the CALICE Collaboration, in particular V.~Boudry, R.~Cornat, D.~Decotigny and F.~Gastaldi, 
and members of the EUDET DAQ group for their support, practical contributions and valuable discussions.  We would also 
like to acknowledge the support of the Science and Technology Facilities Council and the Commission of the European 
Communities under the 6$^{\rm th}$ Framework Programme ``Structuring the European Research Area", Contract number 
RII3-026126.  One of us (M. Wing) acknowledges the support of DESY, Hamburg and the Alexander von Humboldt Stiftung.

\end{document}